# Node Dissimilarity Index for Complex Network Analysis

Natarajan Meghanathan
Professor of Computer Science
Jackson State University, Jackson, MS 39217
Email: natarajan.meghanathan@jsums.edu

**Abstract**
We propose a principal component analysis (PCA)-based approach to quantify (the node dissimilarity index, NDI) the extent of dissimilarity among nodes in a network with respect to values incurred for a suite of node-level metrics (like centrality metrics). We subject the dataset ($n$ nodes and their values incurred for four commonly studied centrality metrics: degree, eigenvector, betweenness and closeness) to PCA and retain the $m$ ($\leq 4$) principal components (with variance $\geq 1.0$). We construct an $n$-node dissimilarity matrix whose entries are the absolute difference (if $m = 1$) or Euclidean distance (if $m > 1$) of the principal component coordinates of the corresponding nodes. We compute NDI ($\geq 1.0$) to be the ratio of the principal Eigenvalue of the node dissimilarity matrix and average of entries in the node dissimilarity matrix. The larger the NDI, the greater the dissimilarity among the node-level metrics (centrality metrics) values considered for analysis.

## 1 Introduction

We explore the problem of quantifying the extent of dissimilarity among the nodes in a complex network with respect to a suite of node-level metrics (such as the centrality metrics [1]). The existing work in the literature to address the above problem focuses on the use of similarity assessment methods and metrics, either at the node-level or at the network-level. The metrics for similarity assessment are typically at the node-level: either pair-wise (like the cosine similarity index [2], matching index [3], etc) or for a set of nodes (such as the Rich club coefficient [4]) and sometimes at the network-level as well (such as the node similarity index [5]). Our focus in this paper is both at network-level and node-level dissimilarity (rather than similarity) assessment and quantification.

We consider a suite of the four commonly studied centrality metrics (such as the degree: DEG [1], eigenvector: EVC [6], betweenness: BWC [7] and closeness: CLC [8]) as the node-level metrics with respect to which we seek to quantify the extent of dissimilarity among the nodes. Nevertheless, the approach proposed in this paper could be used with respect to any node-level metric, including the non-topology based ones (for e.g., node age, node height, etc in social networks). While the degree centrality of a node is the number of neighbors of the node, the eigenvector centrality of a node is a measure of the degree of the node as well as the degrees of its neighbors (and is quantified as the entry for the node in the principal Eigenvector [9] of the 0-1 adjacency matrix of the graph); the betweenness centrality of a node is a measure of the fraction of the shortest paths between any two nodes in the network going through the node (and is quantified as the sum of these fractions) and the closeness centrality of a node is a measure of the proximity of the node to the rest of the nodes in the network (and is measured as the inverse of the sum of the lengths of the shortest paths of the node to the rest of the nodes in the network).

We assume the node-level metrics considered for analysis to exhibit moderate-strong correlation so that we could use principal component analysis (PCA [9]) to reduce the number of dimensions, but at the same time maximally capture the extent of variations in the values incurred by the nodes with respect to these metrics. Several works in the literature (e.g., [10]) have demonstrated moderate-strong correlation among the above four centrality metrics considered for our research. The use of PCA facilitates the approach to be extendable for multi-dimensional datasets.

The proposed PCA-based approach is briefly explained below: We consider a $n$x4 dataset comprising of the centrality metrics values incurred by the $n$ nodes in a complex network with respect to the neighborhood-based DEG and EVC metrics and the shortest paths-based BWC and CLC metrics. We subject the dataset to PCA and retain the $m$ ($m \leq 4$) principal components whose variance is greater than or equal to 1.0. We form a $m$-dimensional coordinate system for the nodes wherein the coordinates of a node are the entries for the node in the $m$ retained principal components. We build a distance matrix (referred to as the Node Dissimilarity Matrix: NDM) whose value for a cell ($i$, $j$) indicates the absolute difference (if $m = 1$) or the Euclidean distance (if $m > 1$) between the entries for the nodes i and j in the $m$ principal components. We propose a metric (both at the network-level and the individual node-level) called the Node Dissimilarity Index (NDI) to seamlessly capture the extent of dissimilarity among the nodes in the network with respect to the node-level metrics (in our case, the four centrality metrics). We propose to measure the network-level NDI as the ratio of the principal Eigenvalue (a measure of the extent of variation in the entries) of the NDM and the average of the entries in the NDM. Since the principal Eigenvalue of a symmetric matrix with positive entries (including 0) is greater than or equal to the average of the entries in the symmetric matrix [11], the network-level NDI value is guaranteed to be greater than or equal to 1.0. The larger the NDI value for a network, the larger the extent of dissimilarity among the nodes in the network with respect to the node-level metric values considered. The network-level NDI metric could be thence used to seamlessly compare two different networks (even with extremely different values for the number of nodes and edges) with respect to the extent of dissimilarity among the nodes in these networks for a suite of node-level metrics. In addition to the network-level NDI metric, we also propose that the principal Eigenvector of the NDM could be used to individually quantify (referred to as node-level NDI) and rank the nodes on the basis of the extent of dissimilarity with respect to the suite of node-level metrics considered. The node-level NDI values are expected to be high for nodes that exhibit significantly lower or significantly higher values for the node-level metrics compared to the rest of the nodes. Such nodes could be branded as outlier nodes and the node-level NDI measure could quantify the extent of outerlierness of the individual nodes.

The rest of the paper is organized as follows: Section 2 presents other related work and highlights the uniqueness of the proposed approach and the NDI metric. Section 3 presents the proposed PCA-based approach to quantify the extent of node dissimilarity along with a running example toy graph. Section 4 presents the network-level NDI values obtained for real-world networks by applying the proposed PCA-approach as well as conducts a correlation study with the unit-disk graph based node similarity index (NSI) values obtained for these networks. Section 4 also visualizes the node-level NDI values for the real-world networks to identify nodes that are the most dissimilar compared to the rest of the nodes. Section 5 concludes the paper and presents plans for future work. Throughout the paper, the terms 'node' and 'vertex', 'edge' and 'link', 'network' and 'graph', 'measure' and 'metric' are used interchangeably. They mean the same.

## 2  Related Work

Node-level similarity measures have been so far proposed at the structural (Jaccard coefficient [9]), attribute (Cosine index [2]) and behavioral-levels (Pearson's correlation [9]). Nevertheless, all these three levels could only be used to assess the extent of (pair-wise) similarity between any two nodes in the network. The Rich club coefficient [4] could be used to assess the extent of similarity among a set of nodes in the network. To the best of our knowledge, there is no node-level similarity or dissimilarity measure proposed to quantify the extent to which a particular node is different from the rest of the nodes in the network with respect to a suite of node-level metrics (such as centrality metrics). Also, at the network-level, the node similarity index (NSI) measure is the only and most related earlier work (to the problem considered in this paper) proposes a measure called the node similarity index (NSI) to quantify the extent of similarity among the nodes in a network with respect to a suite of $k$ node-level metrics. However, the value of the NSI measure (more details below) could be easily swayed towards just a single outlier node that is much different from the rest of the nodes in the network. The NSI measure also needs to build a logical coordinate system of dimensions corresponding to the number of node-level metrics considered (not scalable) for its computation. The proposed network-level and node-level NDI metrics address the above shortcomings (explained below).

The algorithm behind the computation for NSI builds a unit-disk graph on a logical *k*-dimensional coordinate system wherein the entry for a node is the normalized values with respect to the *k* node-level metrics. Two nodes are considered to be connected with an edge in this coordinate system if the Euclidean distance between their coordinates is less than or equal to a threshold distance. The algorithm conducts binary search (on a distance scale of 0 to $\sqrt{k}$) to determine the minimum threshold distance that would result in a connected graph of the nodes in the logical *k*-dimensional coordinate system. The NSI metric is then computed as 1 - (the minimum threshold distance / $\sqrt{k}$).

The NSI metric is more of a reflection of the extent of outlierness among the nodes in a network, rather than the similarity. If there exists one or more outlier nodes (whose node-level metrics/centrality values are much different than the rest of the nodes in the network), the NSI metric value would reflect how far are these outlier nodes vis-a-vis the rest of the nodes in the logical *k*-dimensional coordinate system. Also, there is no dimensionality reduction with the computation procedure for the NSI metric, making it less scalable for high-dimensional datasets. On the other hand, the proposed NDI metric uses PCA for dimensionality reduction and takes a comprehensive approach of building a *n* x *n* node dissimilarity matrix (NDM: whose entries are the pair-wise difference/distance between any two nodes in the coordinate system based on the retained principal components) and is quantified based on the NDM's principal Eigenvalue. The NSI metric is only a network-level metric, whereas the NDI metric can be measured at both the network-level and node-level. The NSI metric could only present a holistic quantification of the extent of outlierness among the nodes in the network, whereas the node-level NDI metric could be used to quantify and rank the individual nodes on the basis of the extent of outlierness.

## 3  PCA-based Approach to Determine the Node Dissimilarity Index

Given a network/graph of 'n' nodes, we determine the values incurred by the nodes with respect to the following four centrality metrics: degree (DEG), eigenvector (EVC), betweenness (BWC) and closeness (CLC). The dataset (n rows and 4 columns; where 4 is the number of centrality metrics considered) used for PCA comprises of the standardized values of the nodes with respect to each of the four centrality metrics. As part of the PCA procedure, we first determine the covariance matrix of the n x 4 dataset. An entry for cell (*p*, *q*) in the covariance matrix (of dimensions 4 x 4) represents the Pearson's correlation coefficient between the centrality metrics corresponding to columns *p* and *q*. We then determine the 4 Eigenvalues and the corresponding 4 Eigenvectors (one Eigenvector for each Eigenvalue) of the covariance matrix. The 4 principal components (PCs) for the dataset are then obtained by multiplying the standardized n x 4 dataset with the 4 x 1 Eigenvectors. We determine the variance of the entries in each principal component (the variance of a PC is its corresponding Eigenvalue divided by *n*-1, where *n* is the number of rows/nodes in the standardized dataset) and retain only those principal components whose variance in the entries is greater than or equal to 1.0. Let 'm' represents the number of such retained principal components. Figure 1 presents an example graph; the raw and standardized values (the latter forming the dataset for PCA) of its centrality metrics; the 4 x 4 covariance matrix of the n x 4 standardized dataset as well as its 4 Eigenvalues and the corresponding Eigenvectors. Figure 1 also presents the computation of the principal component (whose variance is greater than 1.0) that is to be retained for NDI computation. The computation of the other three principal components are not shown as they will not be retained for further analysis (since their variance is less than 1.0, as shown in Figure 1).

We now construct a logical m-dimensional coordinate system wherein the coordinates of a node are the corresponding entries for the node in the 'm' retained principal components of the dataset. We determine a n x n distance matrix (for the m-dimensional coordinate system) referred to as the Node Dissimilarity Matrix (NDM) that captures the Euclidean distance (if m > 1) or the absolute difference (if m = 1) between any two nodes in the network. If two nodes incurred similar values for the centrality metrics (i.e., centrality values close enough to each other), they are expected to be located closer to each other in the logical m-dimensional coordinate system and their entry in the NDM would be smaller. On the other hand, if two nodes are much different (dissimilar) from each other, the entry for the pair in the NDM would be larger. The entries in the NDM would thus be a measure of the extent of dissimilarity among the nodes in the network. We determine the principal Eigenvalue of the NDM as well as the average of the non-diagonal entries (sum of the non-diagonal entries divided by n-1, where n is the number of nodes in the network) in the NDM. The NDI (node dissimilarity index) is the measured the ratio of the principal Eigenvalue of the NDM and the average of its non-zero diagonal entries. For a symmetric matrix with positive entries, the principal Eigenvalue of the matrix is guaranteed to be greater than or equal to the average of the entries. Hence, the NDI value for a network is expected to be greater than or equal to 1.0. Figure 1 presents the NDM for the example graph based on the entries for the nodes in the retained principal component. The largest Eigenvalue for the NDM is 16.5251 and the average of the entries in the NDM is 15.4584. The NDI value for the toy example graph with respect to the four centrality metrics (DEG, EVC, BWC and CLC) is thus 16.5251 / 15.4584 = 1.069.

Figure 2 presents the node-level NDI values that correspond to the individual entries for the nodes in the principal Eigenvector of the NDM. We notice that nodes 4 and 5 (nodes that incurred much larger centrality values) as well as node 0 (node that incurred much lower centrality values) incur relatively larger node-level NDI values, justifying the intent of the metric to quantify the extent of dissimilarity (at the either of the two extremes) compared to the rest of the nodes in the network. The rest of the nodes (nodes 1, 2, 3, 6 and 7) incur similar NDI values that are also much lower than the high NDI values of nodes 0, 4 and 5. We formally confirm and validate this categorization through the use of the well-known Elbow method [12]: nodes 0, 4 and 5 lie in the upper arm, and nodes 1, 2, 3, 6 and 7 lie in the fore arm when the node-level NDI values are plotted in the decreasing order of the values. In Figure 2, we also accordingly color code the nodes as red (dissimilar nodes from the rest of the network) and green (nodes similar to the rest of the network). As seen in this toy example graph, we expect a majority of the nodes in the network (5/8 = 0.625 fraction of nodes) to be similar to each other and only a relatively smaller fraction (3/8 = 0.375) of the nodes to be dissimilar to the rest of the nodes in the network. Such dissimilar nodes could be nodes with high centrality values (as is the case of nodes 4 and 5 in the example graph) or nodes with low centrality values (like node 0 in the example graph).

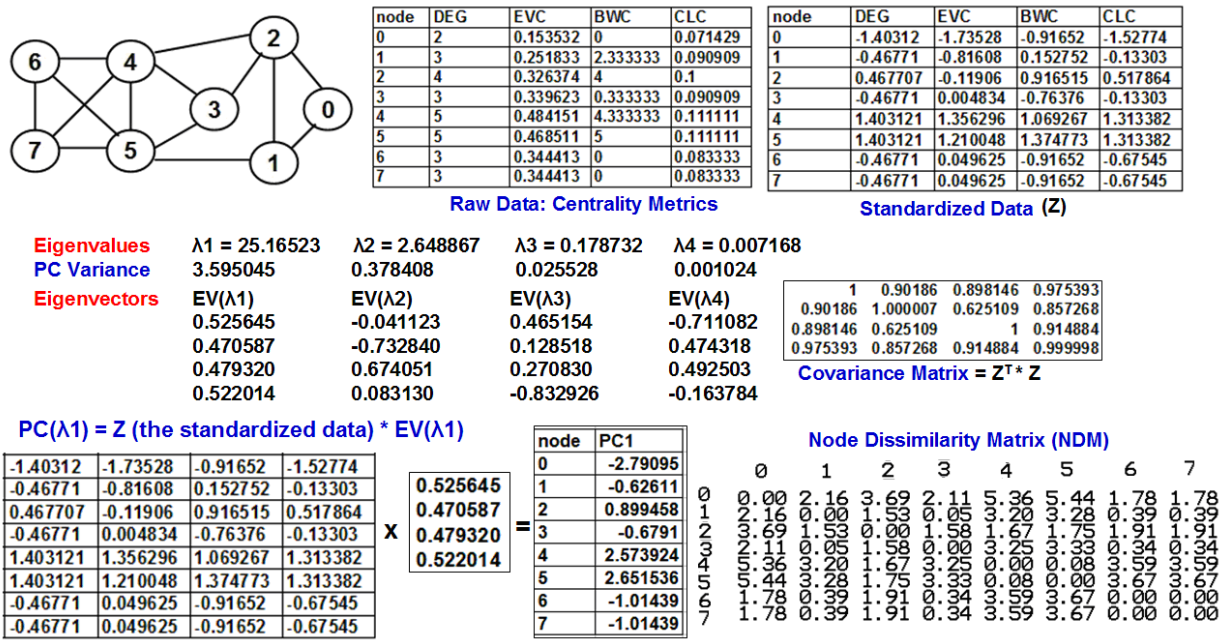

**Figure 1:** Example Graph; its Centrality Metrics; Eigenvalues and Eigenvectors of the Covariance Matrix of the Standardized Dataset; the Principal Component Retained

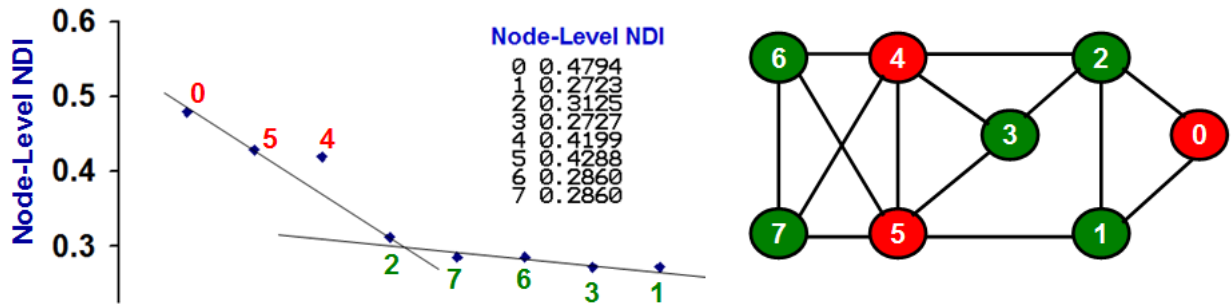

**Figure 2:** Node-Level NDI Values for the Example Graph of Figure 1 and the Use of the Elbow Method to Categorize Nodes (Dissimilar: Red and Similar: Green)

Note that the node-level NDI values incurred by the nodes in a graph range from 0 to 1, as these are the entries in the principal Eigenvector of the NDM and the Eigenvector is normalized. Due to the normalized nature of the Eigenvector, the entries in the Eigenvector are on a relative scale. Hence, the node-level NDI values of the nodes in one real-world network are not comparable to those incurred for the nodes in another real-world network. In other words, the node-level NDI value captures the extent of dissimilarity of a node with the rest of the nodes in the same network, but not a different network.

## 4 Node Dissimilarity Index (NDI) Values for Real-World Networks

We applied the proposed PCA-based approach of Section 2 on the centrality metrics datasets of real-world networks. In this section, we present the network-level and node-level NDI results obtained for a suite of 12 real-world networks, whose spectral radius ratio of node degree [11] ($\lambda_{sp}$; a metric that captures the extent of variation in node degree) ranges from 1.01 to 3.81: i.e., from random networks to scale-free networks. Table 1 presents these 12 real-world networks and their network-level NDI values obtained as well as their unit-disk graph based node similarity index (NSI) values. Figure 3 plots the network-level NDI vs. $\lambda_{sp}$ and NSI metrics. We observe only a moderately positive correlation ($R^2$ for a linear fit: 0.4083) between network-level NDI and $\lambda_{sp}$ and a moderately negative correlation ($R^2$ for a linear fit: 0.5382) between network-level NDI and NSI. This confirms the uniqueness of the network-level NDI metric to comprehensively capture and quantify the extent of dissimilarity among the nodes in a network with respect to a suite of node-level (centrality) metrics.

Table 1: Network-Level NDI and NSI Metrics for a Suite of Real-World Networks

| Net # | Network Name | # Nodes | # Edges | $\lambda_{sp}$ | NDI | NSI |
|---|---|---|---|---|---|---|
| Net 1 | US Football Net | 115 | 613 | 1.01 | 1.0699 | 0.9858 |
| Net 2 | Taro Exchange Net | 22 | 39 | 1.06 | 1.0946 | 0.9333 |
| Net 3 | Flying Teams Cadets Net | 48 | 170 | 1.21 | 1.1324 | 0.9189 |
| Net 4 | Dolphin Net | 62 | 159 | 1.40 | 1.0750 | 0.9000 |
| Net 5 | Band Jazz Net | 198 | 2742 | 1.44 | 1.1631 | 0.7856 |
| Net 6 | Karate Net | 34 | 78 | 1.47 | 1.1966 | 0.8453 |
| Net 7 | Adjacency Noun Net | 112 | 425 | 1.73 | 1.2964 | 0.8077 |
| Net 8 | Les Miserables Net | 77 | 254 | 1.82 | 1.2501 | 0.6793 |
| Net 9 | Copper Field Net | 87 | 406 | 1.83 | 1.5554 | 0.5229 |
| Net 10 | Anna Karenina Net | 138 | 493 | 2.48 | 1.4300 | 0.8163 |
| Net 11 | US Airports 1997 Net | 332 | 2126 | 3.22 | 1.3178 | 0.8892 |
| Net 12 | EU Air Transport Net | 405 | 1981 | 3.81 | 1.3869 | 0.8419 |

Figure 4 presents the distributions of the sorted order of the node-level NDI values for the nodes in 3 of the 12 real-world networks listed in Table 1. The three networks chosen for the plots in Figure 4 incur network-level NDI values that are uniformly distributed in the range [1.0699, ..., 1.5554]. We observe majority of the nodes in each of these networks to incur comparably similar node-level NDI values; only a few nodes incur appreciably larger NDI values (Figure 4 displays the fraction of dissimilar nodes obtained using the Elbow method).

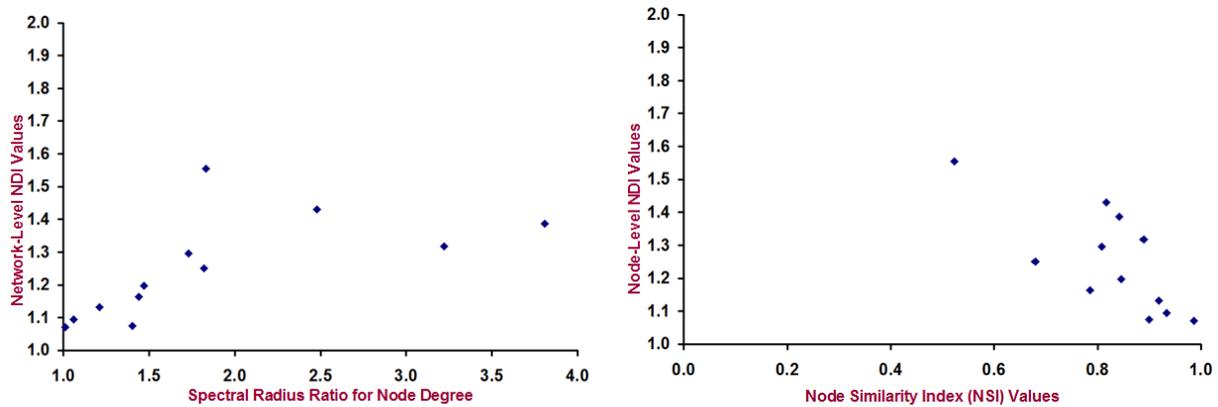

**Figure 3:** Network-Level NDI vs. {Spectral Radius Ratio for Node Degree, Node Similarity Index} for Real-World Networks

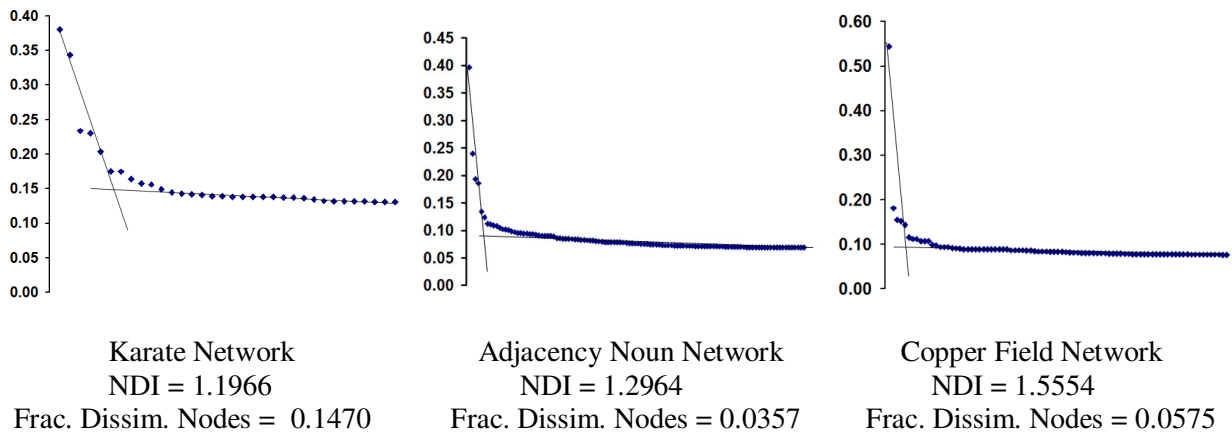

| Karate Network | Adjacency Noun Network | Copper Field Network |
| --- | --- | --- |
| NDI = 1.1966 | NDI = 1.2964 | NDI = 1.5554 |
| Frac. Dissim. Nodes =  0.1470 | Frac. Dissim. Nodes = 0.0357 | Frac. Dissim. Nodes = 0.0575 |

**Figure 4:** Distribution of the Sorted Node-Level NDI Values of the Real-World Networks

## 5  Conclusions and Future Work

The high-level contributions of this paper are as follows: We have proposed the Node Dissimilarity Index (NDI) measure at both the network-level and node-level with respect to a suite of node-level metrics (such as centrality metrics). The NDI measure comprehensively captures the extent of dissimilarity among the nodes in the network owing to its computation of the node dissimilarity matrix (NDM: that has entries for any two nodes in the network) based on the dominating principal components of the node-level metrics dataset and further subjecting the NDM to a second round of principal component analysis (PCA), making it more scalable for high-dimensional datasets. The NDI metric ($\geq 1.0$) is independent of the number of nodes and edges in the network and hence can be seamlessly used to compare the extent of dissimilarity among the nodes for two different networks. We propose to apply the Elbow method to the sorted distribution (from high to low) of the node-level NDI values and categorize nodes to be dissimilar to the rest of the nodes or similar to the rest of the nodes. The dissimilar nodes are nodes whose values or the node-level metrics are either extremely larger or extremely smaller compared to those incurred by the rest of the nodes in the network. We have also shown that the proposed network-level NDI measure is different (through correlation analysis) from some of the existing network-level metrics in the literature.

As part of future work, we plan to investigate the correlation between the network-level NDI metric and the network-level measures such as density, assortativity index [13] and randomness index [14] as well as propose a mathematical model to assess the node-level NDI values using pair-wise node-level similarity metrics (such as Jaccard coefficient, Cosine index and etc).

**Acknowledgments**
The work leading to this paper was partly funded through the U.S. National Science Foundation (NSF) grant OAC-1835439 and partly supported through a sub contract received from University of Virginia titled: Global Pervasive Computational Epidemiology, with the National Science Foundation as the primary funding agency. The views and conclusions contained in this paper are those of the authors and do not represent the official policies, either expressed or implied, of the funding agency.